\documentclass[a4paper,11pt]{article}

\usepackage{jcappub} 

\usepackage[T1]{fontenc} 
\usepackage{amsthm,amsmath,amssymb}
\usepackage{mathrsfs}
\usepackage{graphicx}

\title{Anisotropic Hyperbolic Inflation}

\author{Chong-Bin Chen,}
\author{Jiro Soda}

\affiliation[a]{Department of Physics, Kobe University, Kobe 657-8501, Japan}

\emailAdd{chongbin@stu.kobe-u.ac.jp}
\emailAdd{jiro@phys.sci.kobe-u.ac.jp}

\abstract{Hyperbolic inflation is an extension of the slow-roll inflation in multi-field models. 
We extend hyperbolic inflation by adding
a gauge field and find four-type attractor solutions:  slow-roll inflation, hyperbolic inflation, anisotropic slow roll inflation, and anisotropic hyperbolic inflation. 
We perform the stability analysis with the dynamical system method.
We also study the transition behaviors of solutions between anisotropic slow roll inflation and anisotropic hyperbolic inflation. Our result indicates that destabilization of the standard slow-roll inflation ubiquitously occurs in multi-scalar-gauge field inflationary scenarios. 
}

\begin{document}
\maketitle
\flushbottom
\section{Introduction}
An inflationary scenario proposed in \cite{Guth:1980zm,Sato:1981ds}, and subsequently developed to slow-roll inflation~\cite{Linde:1981mu,Albrecht:1982wi}, not only provides a solution to the problems of the hot big bang theory, but also accounts for the origin of the large-scale structure of the universe. 
Observation of the temperature fluctuations of the cosmic microwave background radiation (CMB)~\cite{Ade:2015lrj,Array:2015xqh} tells us that
primordial fluctuations are statistically  isotropic, scale invariant and Gaussian. 

An inflationary expansion of the universe is driven by a scalar field. 
Roughly speaking, single-scalar-field slow roll  inflationary models with a flat potential have been successful in accounting for observations. 
While precise cosmological observations force us to probe the inflation with greater precision and to discuss the deviation from the statistical isotropy, the scale invariance of the spectrum and the Gaussian statistics. Remarkably, observations suggest the statistical anisotropy in the temperature fluctuations of the cosmic microwave background, which indicates violation of rotation symmetry of the universe during an inflationary period. In other words, there is a preferred direction $\boldsymbol{n}$ in space of universe so that the power spectrum deviates scale invariant: $P(\boldsymbol{k})=P_{iso}(k)\left(1+g(k)(\hat{\boldsymbol{k}}\cdot \hat{\boldsymbol{n}})^2\right)$ \cite{Ackerman:2007nb}, where $g(k)$ is a parameter characterizing
the amplitude of violation of rotational symmetry and $P_{iso}$ is the isotropic part of power spectrum. The current bound of $g$ is given by current Planck data $|g|\lesssim 10^{-2}$ \cite{Ade:2015lrj,Planck:2015igc}.
Naively, one can expect the anisotropy originates from gauge fields although the cosmic no-hair conjecture implies matters like vector fields will be diluted during inflation thus universe always approaches de Sitter spacetime. In fact, it turns out that the no-hair conjecture does not hold in the inflationary models with a gauge kinetic function~\cite{Watanabe:2009ct,Kanno:2009ei}(see also \cite{Soda:2012zm,Maleknejad:2012fw} for reviews). The models are motivated by the bosonic part of the supergravity action, which has non-minimal coupling
\begin{equation}
    S_{\text{gauge}}=
    -\frac{1}{4}\int d^4x\sqrt{-g}f_{ab}(\phi)F^a_{\mu\nu}F^{b\mu\nu}
\end{equation}
with a metric $f_{ab}(\phi)$ in the gauge field space.
The energy density of gauge fields can remain constant for the kinetic function $f_{ab}(\phi)=\delta_{ab}\exp^{2C\int V/V_{,\phi}}$, where $C$ is a constant, thus produces a nondecaying anisotropy during inflation. Remarkably, the anisotropy $\Sigma/H$ of these classes of models is on the order of slow-roll parameter $\epsilon$, where $\Sigma$ is an anisotropic shear of spacetime and $H$ is the Hubble expansion rate. In other words, the gauge field with a gauge kinetic function destabilize the conventional slow roll inflation and leading
to anisotropic inflation~\cite{Watanabe:2009ct}. 

Theoretically,  inflationary scenarios are faced with a challenge. Indeed, it is difficult to realize slow roll inflation in a low-energy effective field theory of an ultraviolet complete theory, e.g. string theory \cite{Vafa:2005ui,Ooguri:2006in,Obied:2018sgi,Agrawal:2018own,Garg:2018reu}. Notice that string theory or supergravity prefer multi-field inflation rather than the single-field one. The action for scalar fields $\phi^a$ reads
\begin{equation}
    S_{\text{scalar}}
    =-\frac{1}{2}\int d^4x\sqrt{-g} \ G_{ab} (\phi)\ \partial^\mu\phi^a \partial_\mu\phi^{b} \ ,
\end{equation}
where $G_{ab}(\phi)$ is a metric  in scalar field space. 
In the multi-field models, there arises the destabilization of the slow roll inflation due to the geometry in scalar field space~\cite{Renaux-Petel:2015mga}.
It indicates a fertile phase space structure of
the multi-scalar-field inflationary models.
One example is the ``spinflation'', which presents a radial field and some other angular fields \cite{Easson:2007dh}. Inflation occurs even with a steep potential because the radial field is acted upon by a centrifugal force thus rolls down slowly to the bottom of potential. However, the field-space of this model is flat thus the angular momentum will redshift away after a few e-folds. To overcome this problem, the so-called ``hyperbolic inflation'', whose field-space is a hyperbolic space rather than a flat one, was proposed recently~\cite{Brown:2017osf} and 
the observational constraints are investigated~\cite{Mizuno:2017idt,Bounakis:2020xaw}. Because of the exponential contribution from hyperbolic geometry to the angular field the angular momentum can remain relevant during inflation. In this case, the radial field $\phi$ has a new attracting solution when $L\ll 1$ and the potential varies slowly,
\begin{equation}
    \dot{\phi}=-3HL \label{ah} \ ,
\end{equation}
 where $L$ is the curvature scale of the hyperbolic space. A generalization to any number of fields and broken rotational symmetry of the potential were also proposed and tested in swampland conditions \cite{Bjorkmo:2019aev}.

For single-field slow-roll inflation,
the solutions will converge to an attractor trajectory. 
We already know that addition of a gauge field can destabilize
the conventional slow roll inflation. 
We also learn that the geometrical structure 
of the scalar field space can destabilize the conventional slow roll inflation.
Hence, it is natural to expect a novel destabilization occurs in the multi-scalar-gauge-field models. Note that gauge fields can be regarded as one-form fields. We can also consider two-form gauge fields whose field strength is $H^a_{\mu\nu\rho}$~\cite{Ohashi:2013mka,Ito:2015sxj}. 
The general kinetic action can be written as
\begin{equation}
    S_{\text{kin}}=
    \int d^4x\sqrt{-g} \left[
    -\frac{1}{2}G_{ab} (\phi)\ \partial^\mu\phi^a \partial_\mu\phi^{b}
    -\frac{1}{4}f_{ab}(\phi)F^a_{\mu\nu}F^{b\mu\nu}
    -\frac{1}{4}h_{ab}(\phi)H^a_{\mu\nu\rho}H^{b \mu\nu\rho} \right]\ ,
\end{equation}
where $h_{ab}$ is a metric of two-form gauge field space.
It is intriguing to study these general models in detail.
In this paper, to make the analysis precise, we study the simplest example:
hyperbolic inflation coupled to a $U(1)$ gauge filed. 
The analysis of general cases will be reported in a separate paper.
The hyperbolic inflationary model contains a massive radial scalar field $\phi$ and a massless angular scalar field $\theta$. Only the scalar field $\phi$ couples to the $U(1)$ gauge fields with a gauge kinetic function. Because of a new degree of freedom of the gauge field, the original hyperbolic inflation will be unstable in some parameter area. 

It is useful to consider exactly solvable models to grasp a whole picture.
There exist analytical inflationary solutions for an exponential potential~\cite{Halliwell:1986ja,Copeland:1997et}. 
These solutions are the so-called scaling solutions which have a constant slow-roll parameter $\epsilon\equiv-\dot{H}/H^2$, i.e., $\dot{\epsilon}=0$.
This is the generalization of de Sitter solution whose parameter and its derivative are both zero $\epsilon=\dot{\epsilon}=0$ and can be attractors that describe a class of inflationary evolutions at late times.
The exponential potential also gives rise to  analytical anisotropic inflationary solutions~\cite{Kanno:2010nr,Yamamoto:2012tq,Lahiri:2016jqv,Ito:2017bnn,Ohashi:2013pca,Holland:2017cza,Do:2020hjf,Do:2021lyf}. In these cases, the gauge kinetic function is also exponential and the anisotropy remains constant.   In multi-field scenarios, the dynamical trajectory in general deviates from that of single-field inflation \cite{Christodoulidis:2019jsx}. Hence it is worth understanding the phase space structure of solutions of multi-scalar-gauge-field scenarios.
We classify fixed points in the system and perform the stability analysis of the fixed points to reveal the phase space structure.

The organization of the paper is as follow. In section \ref{AnisotropicHyperinflation},  we derive analytical solutions for an exponential potential  and  exponential kinetic function of $\phi$, which contain non-zero constant anisotropy. 
In section \ref{InflationaryAttractorsAnalysis}, we investigate parameter space of these solutions with dynamical system approach and show the attracting behaviors in the anisotropic parameter areas. We also analyze the scaling solutions and anisotropy beyond the exact exponential potential. The final section is devoted to the conclusion.

\section{Anisotropic Hyperbolic Inflation}\label{AnisotropicHyperinflation}
\subsection{Background Equations of Motion}
In this section, we consider the hyperbolic inflationary model coupled with a $U(1)$ gauge field non-minimally. We consider the exponential potential and exponential kinetic function and then find exact anisotropic  power-law solutions in the $\phi\gg L$ regime.

The model is described by the following action
\begin{equation}
    S=\int d^4x\sqrt{-g}\left[\frac{M^2_{pl}}{2}R-\frac{1}{2}G_{ab}g^{\mu\nu}\partial_\mu\phi^a\partial_\nu\phi^b-V-\frac{1}{4}f^2F_{\mu\nu}F^{\mu\nu}\right], \label{action}
\end{equation}
Where $g$ is the determinant of the metric, $R$ is the scalar curvature, $G_{ab}(\phi^c)$ is the metric of the scalar manifold spanned by the fields $\phi^c$ and $V(\phi^c)$ is the potential of scalar fields. The scalar fields couple with gauge fields with a kinetic function $f(\phi^c)$. Here we only consider the $U(1)$ gauge field  $F_{\mu\nu}=\partial_\mu A_\nu-\partial_\nu A_\mu$. We also set $M_{pl}=1$ for the rest of our discussion.

Choosing the gauge $A_0=0$, we can take $x$-axis in the direction of the vector field without loss of generality. Hence the homogeneous gauge field has the form $A_{\mu}=(0,v(t),0,0)$. We also consider the homogeneous scalar fields $\phi^c(t)$. Then we take the anisotropic metric which is only rotational invariant in $y$-$z$ plane as
\begin{equation}
    ds^2=-dt^2+e^{2\alpha(t)}\left[e^{-4\sigma(t)}dx^2+e^{2\sigma(t)}\left(dy^2+dz^2\right)\right],\label{metric}
\end{equation}
where $\alpha(t)$ is an isotropic scale factor and $\sigma(t)$ is a deviation from the isotropy. With these ansatzes, the equation of motion of the gauge field reduces to
\begin{equation}
    \frac{d}{dt}\left[f^2e^{\alpha+4\sigma}\dot{v}\right]=0,
\end{equation}
which can be easily solved to give
\begin{equation}
    \dot{v}=f^{-2}e^{-\alpha-4\sigma}p_A,\label{gaugef}
\end{equation}
where $p_A$ is a constant of integration.

In addition, because of the not-trivial field space $G_{ab}$ of scalar fields,
the equations of motion are given by
\begin{equation}
    \mathcal{D}_t\dot{\phi}^a+3H\dot{\phi}^a+G^{ab}V_{,b}+\frac{1}{2}G^{ab}f_{,b}fF_{\mu\nu}F^{\mu\nu}=0,\label{Phi_q}
\end{equation}
where the covariant directional derivative $\mathcal{D}_t$ is defined by $\mathcal{D}_t X^a=\dot{X}^a+\Gamma^a_{\ bc}\dot{\phi}^b X^c$ for any field space vector, $V_{,b}=\partial V/\partial\phi^b$ and $f_{,b}=\partial f/\partial\phi^b$. We are interested in the two-dimensional hyperbolic field space which has a “radial” field $\phi$ and a "angular" field $\theta$
\begin{equation}
    ds_G^2=d\phi^2+L^2\sinh^2{(\phi/L)}d\theta^2,\label{Smetric}
\end{equation}
where $L$ represents the curvature scale of the field space. 
After substituting (\ref{metric}), (\ref{gaugef}) and (\ref{Smetric}) into the action (\ref{action}) and (\ref{Phi_q}), we obtain the equations of motion
\begin{eqnarray}\label{eom}
    3\dot{\alpha}^2&=&3\dot{\sigma}^2+\frac{1}{2}\dot{\phi}^2+\frac{1}{2}L^2\sinh^2{(\phi/L)}\dot{\theta}^2+V+\frac{p_A^2}{2}f^{-2}e^{-4\alpha-4\sigma},\label{hc}\\
    \ddot{\alpha}&=&-3\dot{\sigma}^2-\frac{1}{2}\dot{\phi}^2-\frac{1}{2}L^2\sinh^2{(\phi/L)}\dot{\theta}^2-\frac{p_A^2}{3}f^{-2}e^{-4\alpha-4\sigma},\label{alpha_q}\\
    \ddot{\sigma}&=&-3\dot{\alpha}\dot{\sigma}+\frac{p_A^2}{3}f^{-2}e^{-4\alpha-4\sigma},\label{sigma_q}\\
    \ddot{\phi}&=&-3\dot{\alpha}\dot{\phi}-V_{,\phi}+L\sinh{(\phi/L)}\cosh{(\phi/L)}\dot{\theta}^2+p_A^2f^{-3}f_{,\phi}e^{-4\alpha-4\sigma},\label{phi_q}\\
    \ddot{\theta}&=&-3\dot{\alpha}\dot{\theta}-\frac{2}{L}\coth{(\phi/L)}\dot{\theta}\dot{\phi},\label{theta_q}
\end{eqnarray}
where $V_{,\phi}=\partial V/\partial\phi$ and $f_{,\phi}=\partial f/\partial\phi$. Here we have assummed the shift symmetry of the angular field $\theta$, i.e., the model is invariant after a shift $\theta\rightarrow\theta+a$ in the field-space. Hence the coupling $f$ is only depended on the radial field $\theta$. We will use these equations of motion to look for isotropic and anisotropic power-law solutions in the regime $\phi\gg L$, where  hyperbolic inflation can occur. 

\subsection{Power-law Solutions of Anisotropic Hyperbolic Inflation}
The anisotropic inflation can be realized for the following exponential potential and kinetic function \cite{Kanno:2010nr}
\begin{equation}
    V(\phi)=V_0e^{\lambda\phi},\ \ \ \ \ \ f(\phi)=f_0e^{\rho\phi} .
\end{equation}
We consider the  regime $\phi\gg L$, where we can use the approximation
\begin{equation}
    \sinh{(\phi/L)}\simeq\cosh{(\phi/L)}\simeq \frac{e^{\phi/L}}{2} \ .\label{approximation}
\end{equation}
Under this approximation, we can find power-law solutions by assuming ansatzes
\begin{equation}
    \alpha=\zeta\log t,\ \ \ \sigma=\eta\log t,\ \ \ \phi=\xi\log t+\phi_0,\ \ \ \theta=\gamma t^p+\theta_0,
\end{equation}
where $\theta_0$ and $\phi_0$ are the initial value of the scalar fields.

For a trivial gauge field, namely,  $p_A=0$, we can obtain the isotropic power-law solution $\eta=0$. One thing worth reminding is that the expansion $\zeta$ is completely derived from equation of the scalar field $\theta$ and only related with $p$ and $\xi$. This is because unlike the original slow-roll inflation, hyperbolic inflation is driven by the ''centrifugal force'' in the hyperbolic space so the potential energy is converted into angular momentum rather than the kinetic energy of $\phi$. From (\ref{hc}) and (\ref{theta_q}), we immediately have 
\begin{equation}
    \zeta=\frac{1}{3}\left(\frac{2}{L\lambda}+1\right),\ \ \ \ \ \xi=-\frac{2}{\lambda},\ \ \ \ \ p=\frac{2}{L\lambda}.\label{pls}
\end{equation}
In order to have a sufficiently fast expansion, we need $L\lambda\ll 1$. 
Substituting these solutions into other equations of motion, we have 
\begin{equation}
    \gamma^2e^{\phi_0/L}=2\lambda\left(\frac{2+L\lambda}{3L}-\frac{2}{\lambda}\right),\ \ \ \ V_0e^{2\lambda\phi_0}=\frac{2}{3L\lambda}\left(\frac{2}{L\lambda}+1\right).
\end{equation}
The isotropic solutions without the gauge field and their stability have already been studied \cite{Mizuno:2017idt, Christodoulidis:2019jsx}.

Next, let us find anisotropic power-law solutions which we are more interested in. From the hamiltonian constraint equation (\ref{hc}), we have relations
\begin{equation}
    \lambda\xi=-2,\ \ \ \ \ \rho\xi+2\zeta+2\eta=1,\ \ \ \ \ \frac{2\xi}{L}+2p-2=-2
\end{equation}
to have the same time dependence for each term. Because the scalar field $\theta$ is decoupled with the gauge field, we have the same solution of $\zeta$ from (\ref{theta_q}). Therefore from (\ref{hc}) and (\ref{sigma_q}), we immediately have the same solutions as isotropic case (\ref{pls}) and non-zero anisotropy 
\begin{equation}
    \eta=\frac{1}{6}+\frac{\rho}{\lambda}-\frac{2}{3L\lambda}.\label{eta_s}
\end{equation}
Then for the amplitudes to balance in (\ref{sigma_q}), we obtain 
\begin{equation}
    w\equiv p_A^2f_0^{-2}e^{-2\rho\phi_0}=\frac{6}{L\lambda}\left(\frac{1}{6}+\frac{\rho}{\lambda}-\frac{2}{3L\lambda}\right).\label{w_s}
\end{equation}
Similarly, for the amplitudes to balance in (\ref{hc}) and (\ref{alpha_q}), we need
\begin{eqnarray}
    3\zeta^2&=&3\eta^2+\frac{1}{2}\xi^2+\frac{1}{2}L^2u_L^2\gamma^2p^2+u_{\lambda}+\frac{1}{2}w,\\
    \zeta&=&3\eta^2+\frac{1}{2}\xi^2+\frac{1}{2}L^2u_L^2\gamma^2p^2+\frac{1}{3}w,
\end{eqnarray}
where we have defined
\begin{equation}
    u_L=\frac{1}{2}e^{\phi_0/L},\ \ \ \ \ \ u_{\lambda}=V_0e^{2\lambda\phi_0}.
\end{equation}
Substituting (\ref{pls}), (\ref{eta_s}) and (\ref{w_s}) into the above two equations, we obtain
\begin{equation}
    u_{\lambda}=\frac{1}{L\lambda}\left(\frac{2}{L\lambda}-\frac{\rho}{\lambda}+\frac{1}{2}\right)
\end{equation}
and
\begin{equation}
    u_L^2\gamma^2=\frac{\lambda^2}{2}\left(\frac{1}{L\lambda}-\frac{2}{\lambda^2}-\frac{\rho}{\lambda}-\frac{3\rho^2}{\lambda^2}+\frac{2\rho}{L\lambda^2}+\frac{1}{4} \right).\label{ulg}
\end{equation}
We do not need to solve $u_L$ and $\gamma$ respectively because $u_L$ is the initial position of $\phi$ and $\gamma$ is the initial angular velocity. Hence $u_L^2\gamma^2=$ const. only reflects the conservation of angular energy. The parameter $\gamma$ has two different values (the positive one and the negative one). 
Note that we have solved all the parameters and (\ref{phi_q}) is automatically satisfied. It is easy to see this solution satisfies hyperbolic inflation attractor (\ref{ah}) if $L\lambda\ll 1$.

\begin{figure}[tbp]
\centering
\includegraphics[scale=0.60]{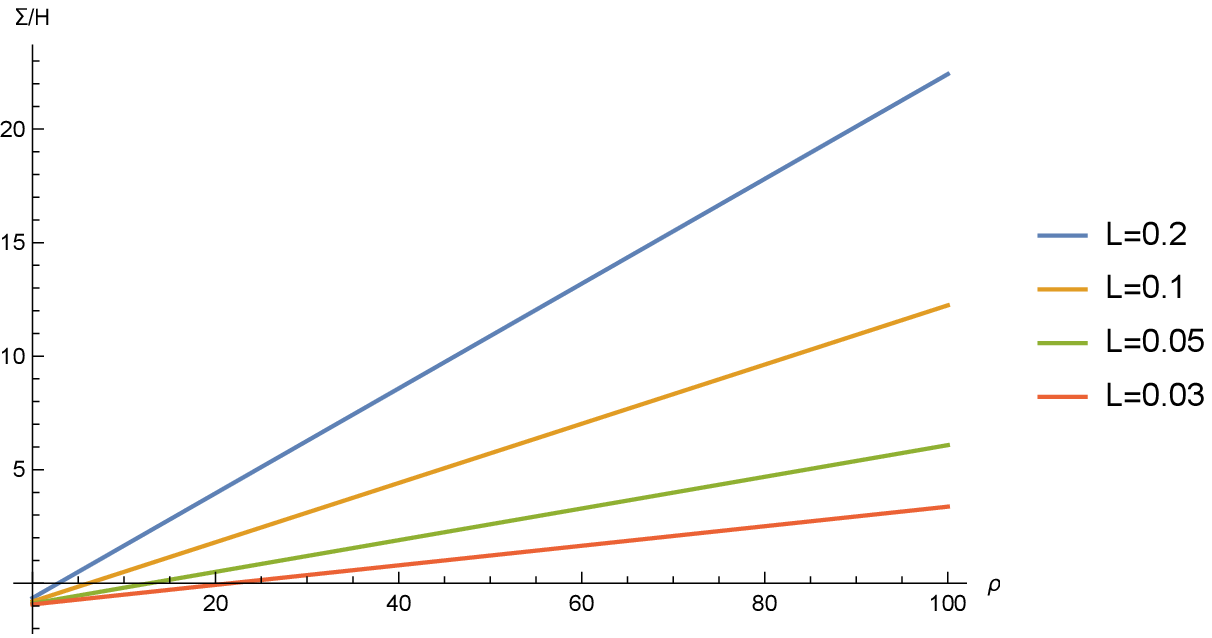}
\hspace{0.1in}
\includegraphics[scale=0.60]{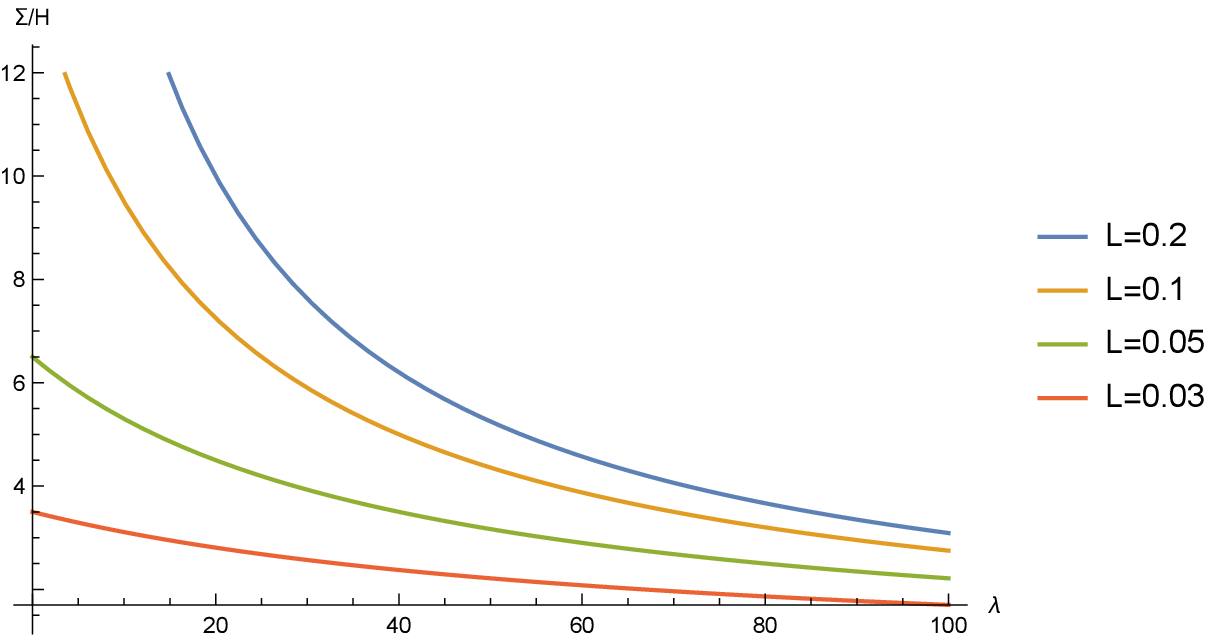}
\caption{\label{fig:ES} The Anisotropy $\Sigma/H$ varies with (left) $\rho$(for $\lambda=3$) and (right) $\lambda$(for $\rho=100$) for different values of $L$ shown.}
\end{figure}

Since the definition $w$ and $u_{\lambda}$ should be positive, we have inequalities of the parameter $\rho$ and $\lambda$
\begin{equation}
    \frac{2}{3L\lambda}-\frac{1}{6}<\frac{\rho}{\lambda}<\frac{2}{L\lambda}+\frac{1}{2}. \label{rho_ineq}
\end{equation}
Under the  condition for inflation $L\lambda\ll 1$, these inequalities reduce to $2/3L<\rho<2/L$. The r.h.s of (\ref{ulg}) should be positive, so we have
\begin{equation}
    4\lambda-8L-4\rho L\lambda-12\rho^2 L+8\rho+L\lambda^2>0.
\end{equation}
Since $L$ is the scale in the denominator of an irrelevant operator in the effective field theory, we expect $L$ to be of the order of the UV cutoff, i.e., $L\ll 1$ \cite{Brown:2017osf}. Hence we need $\rho\gg 1$. The inflation also implies $L\lambda\ll 1$, thus $\rho\gg \lambda$.

The average slow-roll parameter $\epsilon$ in terms of the Hubble parameter $H=\dot{\alpha}$ reads
\begin{equation}
    \epsilon\equiv -\frac{\dot{H}}{H^2}=\frac{1}{\zeta}=\frac{3L\lambda}{2+L\lambda}.
\end{equation}
In the inflationary limit, it reduces to $\epsilon=3L\lambda/2\ll 1$. The anisotropy is characterized by
the ratio
\begin{equation}
    \frac{\Sigma}{H}\equiv \frac{\dot{\sigma}}{\dot{\alpha}}=\frac{\eta}{\zeta}=\frac{1}{3}I\epsilon \ , 
\end{equation}
where
\begin{equation}
    I=\frac{c-1}{c}, \ \ \ \ \ c=\frac{2L\lambda}{L\lambda-6L\rho+4}.
\end{equation}
From the inequality (\ref{rho_ineq}), we immediately see $c>1$. For a $c\sim O(1)$ the anisotropy is of the order of the slow-roll parameter and give rise to counter examples to the cosmic no-hair conjecture \cite{Wald:1983ky,Soda:2014awa}. In figure \ref{fig:ES}, we depicted the anisotropy $\Sigma/H$ as a function of $\lambda$
and $\rho$ for various curvature scales $L$. We see the anisotropy is suppressed for small $L$.

\begin{figure}[tbp]
\centering
\includegraphics[scale=0.55]{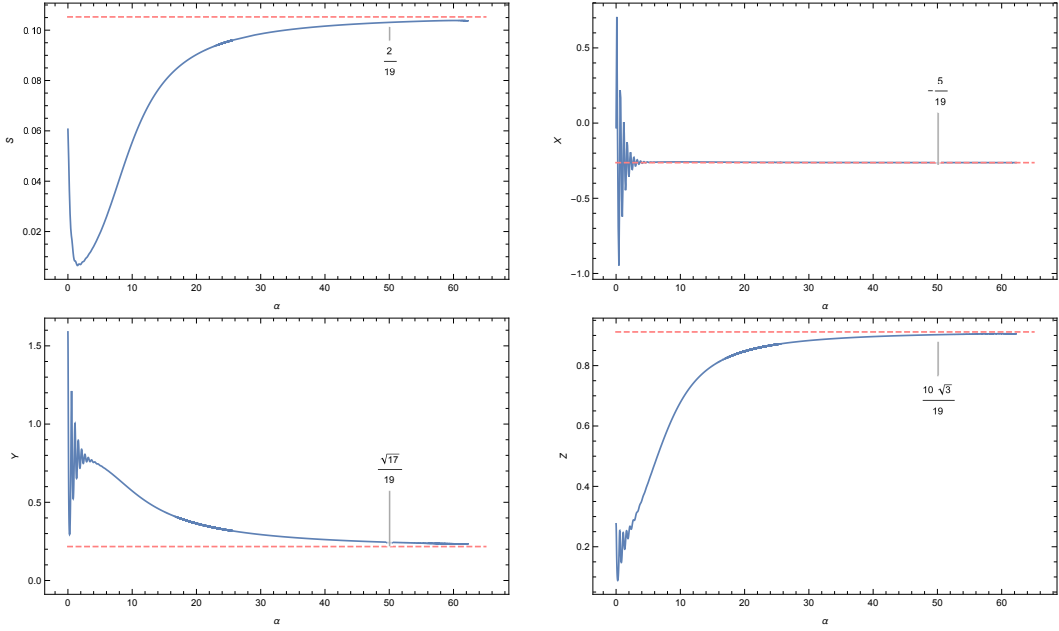}
\caption{\label{Comparision} Evolution of inflationary solutions (solid blue curves) under exponential field space ($\sim e^{\phi/L}/2$) for parameters $\{\lambda,\rho,L\}=\{2.8,7,0.1\}$. The initial conditions we choose are $\phi_0\simeq 23.08$, $\dot{\phi}_0=-0.10$, $\theta_0=0$, $\dot{\theta}_0\simeq 6.36\times 10^{-99}$, $\sigma_0=0.50$,  $\dot{\sigma}_0=0.20$, $p_A\simeq 3.51\times 10^{70}$, $V_0\simeq 1.60\times 10^{-27}$ and $f_0=1.00$. These solution converge to the scaling solutions of anisotropic hyperbolic inflation (dashed red curves).}
\end{figure}

In figure \ref{Comparision}, we numerically solve the equations of motion (\ref{eom})-(\ref{theta_q}) under
the approximation (\ref{approximation}) and show the attractor behaviors of four variables defined by (\ref{xyzs}). These four variables finally converge to the scaling solutions (dashed red curves). $S$ is the anisotropy $\Sigma/H$ and approaches to a non-zero constant. We will analyze these scaling solutions 
in an entire parameter space of $(\lambda,\rho)$ in the next section.

\section{Inflationary Attractors}\label{InflationaryAttractorsAnalysis}
\subsection{Stability of Inflationary Fixed Points}
In the limit $\phi/L\gg 1$, we found exact anisotropic power-law solutions. Such solutions with a constant parameter $\epsilon$ is called scaling solutions. The scaling attractors in multi-field inflation have been studied in \cite{Christodoulidis:2019jsx} and here we extend it to the case coupled with a gauge field.

It is convenient to define dimensionless variables
\begin{equation}
    S=\frac{\dot{\sigma}}{\dot{\alpha}},\ \ \ \ X=\frac{\dot{\phi}}{\dot{\alpha}},\ \ \ \ Y=\frac{L}{2}e^{\phi/L}\frac{\dot{\theta}}{\dot{\alpha}},\ \ \ \ Z=f e^{-\alpha+2\sigma}\frac{\dot{v}}{\dot{\alpha}}.\label{xyzs}
\end{equation}
With these definitions, from (\ref{alpha_q}) the parameter $\epsilon$ can be written as
\begin{equation}
    \epsilon=\frac{1}{2}X^2+\frac{1}{2}Y^2+3S^2+\frac{1}{3}Z^2.
\end{equation}
For general potential and kinetic functions, we can define driving forces
\begin{equation}\label{pd}
    \lambda_a\equiv \frac{d(\ln V)}{d\phi^a},\ \ \ \ \ \rho_a\equiv\frac{d(\ln f)}{d\phi^a}.
\end{equation}
The hamiltonian constraint equation can be written as
\begin{equation}
    -\frac{V}{\dot{\alpha}^2}=3(S^2-1)+\frac{1}{2}X^2+\frac{1}{2}Y^2+\frac{1}{2}Z^2.
\end{equation}
For a positive potential, we have the inequality
\begin{equation}
    3(S^2-1)+\frac{1}{2}X^2+\frac{1}{2}Y^2+\frac{1}{2}Z^2<0.
\end{equation}
Let us use an e-folding number $\alpha$ as a time. Using the hamiltonian constraint,
we can write the equations of motion in terms of  (\ref{xyzs})  as
\begin{eqnarray}\label{LinearEQ}
    \frac{dS}{d\alpha}&=&\left(\frac{1}{2}X^2+\frac{1}{2}Y^2+3S^2+\frac{1}{3}Z^2-3\right)S+\frac{1}{3}Z^2.\label{S_q},\label{LinearEQS}\\
    \frac{dX}{d\alpha}&=&\left(\lambda_{\phi}+X\right)\left(\frac{1}{2}X^2+\frac{1}{2}Y^2+3S^2+\frac{1}{3}Z^2-3\right)+\frac{1}{L}Y^2+\left(\frac{\lambda_{\phi}}{6}+\rho_{\phi}\right)Z^2,\label{X_q}\label{LinearEQX}\\
    \frac{dY}{d\alpha}&=&\left(\frac{1}{2}X^2+\frac{1}{2}Y^2+3S^2+\frac{1}{3}Z^2-3-\frac{1}{L}X\right)Y,\label{Y_q}\label{LinearEQY}\\
    \frac{dZ}{d\alpha}&=&\left(\frac{1}{2}X^2+\frac{1}{2}Y^2+3S^2+\frac{1}{3}Z^2-2-2S-\rho_{\phi}X\right)Z,\label{LinearEQZ}.
\end{eqnarray}

The scaling solution with constant $\epsilon$ means $X$, $Y$, $Z$ and $S$ stay at a fixed point in the phase space. The fixed point is determined by $dX/d{\alpha}=dY/d{\alpha}=dZ/d{\alpha}=dS/d{\alpha}=0$. Furthermore, to analyze the linear stability of these fixed points, we obtain the local Lyapunov exponents of the Jacobian matrix evaluated by $d\delta X^a/d\alpha=J^a_{\ b}\delta X^b$. These linearized equations for equations of $S$, $X$, $Y$ and $Z$ are given by
\begin{eqnarray}
    \frac{d\delta S}{d\alpha}&=&XS\delta X+YS\delta Y+\frac{2}{3}\left(1+S\right)Z\delta Z \nonumber\\
    &\ &+\left(\frac{1}{2}X^2+\frac{1}{2}Y^2+9S^2+\frac{1}{3}Z^2-3\right)\delta S,\\
    \frac{d\delta X}{d\alpha}&=&\left(X+\lambda_{\phi}+\frac{2}{L}\right)Y\delta Y+\left(\frac{2}{3}X+\lambda_{\phi}+2\rho_{\phi}\right)Z\delta Z+\left(6X+6\lambda_{\phi}\right)S\delta S \nonumber\\
    &\ &+\left(\frac{3}{2}X^2+\frac{1}{2}Y^2+3S^2+\frac{1}{3}Z^2+\lambda_{\phi}X-3\right)\delta X\\
    \frac{d\delta Y}{d\alpha}&=&\left(X-\frac{1}{L}\right)Y\delta X+\frac{2}{3}YZ\delta Z+6YS\delta S \nonumber\\
    &\ &+\left(\frac{1}{2}X^2+\frac{3}{2}Y^2+3S^2+\frac{1}{3}Z^2-3-\frac{1}{L}X\right)\delta Y,\\
    \frac{d\delta Z}{d\alpha}&=&\left(X-\rho_{\phi}\right)Z\delta X+YZ\delta Y+\left(6S-2\right)Z\delta S \nonumber\\
    &\ &+\left(\frac{1}{2}X^2+\frac{1}{2}Y^2+3S^2+Z^2-2-2S-\rho_{\phi}X\right)\delta Z.
\end{eqnarray}

There are four kinds of attractors in this dynamical system, corresponding to four kinds of scaling solutions.
\begin{itemize}
\item[$\bullet$] The isotropic slow roll solution: $Y=S=0$ and $X\neq 0$. In this case, from (\ref{S_q}) we immediately have $Z=0$. Then the equation (\ref{X_q}) yields $X=-\lambda_{\phi}$ or $X=\pm\sqrt{6}$. The latter one corresponds to a configuration whose kinetic term dominates during the inflationary period. We do not consider this case here. Hence the isotropic fixed point is given by
\begin{equation}
    S=0,\ \ \ \ \ X=-\lambda_{\phi},\ \ \ \ \ Y=0,\ \ \ \ \ Z=0.
\end{equation}
The eigenvalues of the Jacobian matrix are calculated as
\begin{eqnarray}
    \omega_1&=&\frac{1}{2}\lambda_{\phi}^2-3,\ \ \ \ \ \  \omega_2=\frac{1}{2}\lambda_{\phi}^2-3,\nonumber\\ \omega_3&=&\frac{L\lambda_{\phi}^2+2\lambda_{\phi}-6L}{2L},\ \ \ \ \ \ \omega_4=\frac{1}{2}\lambda_{\phi}^2-2+\rho_{\phi}\lambda_{\phi}.
\end{eqnarray}
We need all of the eigenvalues to be negative to make the fixed points stable. For a fixed point with $\lambda_{\phi}<\sqrt{6}$, we have negative $\omega_1$ and $\omega_2$. If $L\lambda_{\phi}^2+2\lambda_{\phi}-6L<0$, we have a negative $\omega_3$ 
implying that angular velocity quickly decays. Moreover, $\omega_4$ becomes negative when $\lambda_{\phi}^2+2\rho_{\phi}\lambda_{\phi}-4<0$.

\item[$\bullet$] The anisotropic slow roll solution: $Y=0$ and $X,Y,Z\neq 0$. In this case, there exists a solution $S=2$ which should be excluded. The fixed points corresponding to desired anisotropic solutions are \cite{Kanno:2010nr}
\begin{eqnarray}\label{AG_FP}
    S&=&\frac{2\left(\lambda_{\phi}^2+2\rho_{\phi}\lambda_{\phi}-4\right)}{\lambda_{\phi}^2+8\rho_{\phi}\lambda_{\phi}+12\rho_{\phi}^2+8},\ \ \ \  X=-\frac{12\left(\lambda_{\phi}+2\rho_{\phi}\right)}{\lambda_{\phi}^2+8\rho_{\phi}\lambda_{\phi}+12\rho_{\phi}^2+8},\nonumber\\
    Y&=&0, \ \ \ \ Z^2=\frac{18\left(\lambda_{\phi}^2+2\rho_{\phi}\lambda_{\phi}-4\right)\left(-\lambda_{\phi}^2+4\rho_{\phi}\lambda_{\phi}+12\rho_{\phi}^2+8\right)}{\left(\lambda_{\phi}^2+8\rho_{\phi}\lambda_{\phi}+12\rho_{\phi}^2+8\right)^2}.
\end{eqnarray}
The fixed point requires $\lambda_{\phi}^2+2\rho_{\phi}\lambda_{\phi}-4>0$. For one set of parameters $\{\lambda_{\phi}, \rho_{\phi}, L\}$ there are two different fixed points $(S, X, 0, \pm|Z|)$. Again since we are considering the inflation solution $\lambda_{\phi}\ll 1$, which implies $\rho_{\phi}\gg 1$, the eigenvalues of the Jacobian matrix are given by
\begin{eqnarray}
    \omega_1&\simeq& -3,\ \ \ \ \ \ \ \ \ \omega_{2,4}\simeq -\frac{3}{2}\pm\sqrt{\frac{9}{4}-3\left(\lambda_{\phi}^2+2\rho_{\phi}\lambda_{\phi}-4\right)},\nonumber\\
    \omega_3&=&\frac{3\left(4\lambda_{\phi}-8L-4\rho_{\phi} L\lambda_{\phi}-12\rho_{\phi}^2 L+8\rho_{\phi}+L\lambda_{\phi}^2\right)}{L\left(\lambda_{\phi}^2+8\rho_{\phi}\lambda_{\phi}+12\rho_{\phi}^2+8\right)}.
\end{eqnarray}
Note that here we give an exact $\omega_3$ because it is related to the unstable condition of $\theta$. Hence the fixed point of anisotropic solution is stable for parameters satisfying $\lambda_{\phi}^2+2\rho_{\phi}\lambda_{\phi}-4>0$ and $4\lambda_{\phi}-8L-4\rho_{\phi} L\lambda_{\phi}-12\rho_{\phi}^2 L+8\rho_{\phi}+L\lambda_{\phi}^2<0$.

\item[$\bullet$] The hyperbolic isotropic solution: $S=0$ and $X,Y\neq 0$. When $L\lambda_{\phi}^2+2\lambda_{\phi}-6L>0$ the isotropic slow roll  fixed point becomes unstable and exits from the slow-roll inflation. However, the hyperbolic inflation with small $L\lambda_{\phi}\ll 1$ makes inflation possible again. In the isotropic case we also have $Z=0$. Then the corresponding fixed points are
\begin{equation}
    S=0,\ \ \ \ \ X=-\frac{6L}{L\lambda_{\phi}+2},\ \ \ \ \ Y^2=\frac{6L(L\lambda_{\phi}^2+2\lambda_{\phi}-6L)}{\left(L\lambda_{\phi}+2\right)^2},\ \ \ \ \ Z=0.
\end{equation}
For one set of parameters $\{\lambda_{\phi}, \rho_{\phi}, L\}$ there are two different fixed points $(0, X, \pm|Y|, 0)$. The eigenvalues of the Jacobian matrix of this fixed point in the limit $L\lambda_{\phi}\ll 1$ are given by
\begin{eqnarray}
    \omega_1&\simeq&-3,\ \ \ \ \ \ \ \ \omega_{2,3}\simeq -\frac{3}{2}\mp\frac{3}{2}\sqrt{1-\frac{8}{3}\left(\lambda_{\phi}^2+\frac{\lambda_{\phi}}{L}-3\right)},\nonumber\\
    \omega_4&=&\frac{L\lambda_{\phi}+6L\rho_{\phi}-4}{L\lambda_{\phi}+2}.
\end{eqnarray}
We give an exact $\omega_4$ because it is related to the stability of the solutions of gauge field perturbation $\delta Z$. We can see $\omega_2$ is always negative and $\omega_4<0$ when  $L\lambda_{\phi}+6L\rho_{\phi}-4<0$ so that the isotropy is stable. On the other hand, when $L\lambda_{\phi}\ll 1$, the condition $L\lambda_{\phi}+2\lambda_{\phi}^2-6L>0$ reduces to $\lambda_{\phi}>3L$. The second term in $\omega_3$ is either less than $3/2$ or an imaginary number. Hence the real part of  $\omega_3$ is always negative and the fixed point is stable \cite{Bounakis:2020xaw}.

\item[$\bullet$] The hyperbolic anisotropic solution: $S$, $X$, $Y$, $Z\neq 0$. When  $L\lambda_{\phi}+6L\rho_{\phi}-4>0$ the perturbation of gauge field of hyperbolic inflation becomes unstable. In this case we have anisotropic fixed points of 
hyperbolic inflation,
\begin{eqnarray}\label{AH_FP}
    S&=&\frac{L\lambda_{\phi}+6L\rho_{\phi}-4}{2\left(L\lambda_{\phi}+2\right)},\ \ \ \ \ \ \ \ \ \ \ X=-\frac{6L}{L\lambda_{\phi}+2},\nonumber\\
    Y^2&=&\frac{9L\left(4\lambda_{\phi}-8L-4\rho_{\phi} L\lambda_{\phi}-12\rho_{\phi}^2 L+8\rho_{\phi}+L\lambda_{\phi}^2\right)}{2\left(L\lambda_{\phi}+2\right)^2},\nonumber\\ Z^2&=&\frac{9\left(L\lambda_{\phi}+6L\rho_{\phi}-4\right)}{\left(L\lambda_{\phi}+2\right)^2}.
\end{eqnarray}
Hence we have $4\lambda_{\phi}-8L-4\rho_{\phi} L\lambda_{\phi}-12\rho_{\phi}^2 L+8\rho_{\phi}+L\lambda_{\phi}^2>0$. 
We found for one set of parameters $\{\lambda_{\phi}, \rho_{\phi}, L\}$ there are four different fixed points, which are $(S, X,\pm|Y|,\pm|Z|)$. The eigenvalues of the Jacabian matrix of these fixed points are given by
\begin{eqnarray}
    \omega_{1,2}&=&-\frac{3}{\lambda_{\phi}  L+2}\mp\frac{3\sqrt{A(\lambda_{\phi},\rho_{\phi})}}{2 L (\lambda_{\phi}  L+2)^2},\ \ \   \omega_{3,4}=-\frac{3}{\lambda_{\phi}  L+2}\mp\frac{3\sqrt{B(\lambda_{\phi},\rho_{\phi})}}{2 L (\lambda_{\phi}  L+2)^2},
\end{eqnarray}
where $A$ and $B$ are given by (\ref{eigenvalueA}) and (\ref{eigenvalueB}) respectively.

\end{itemize}
\begin{figure}
\centering
\includegraphics[scale=0.6]{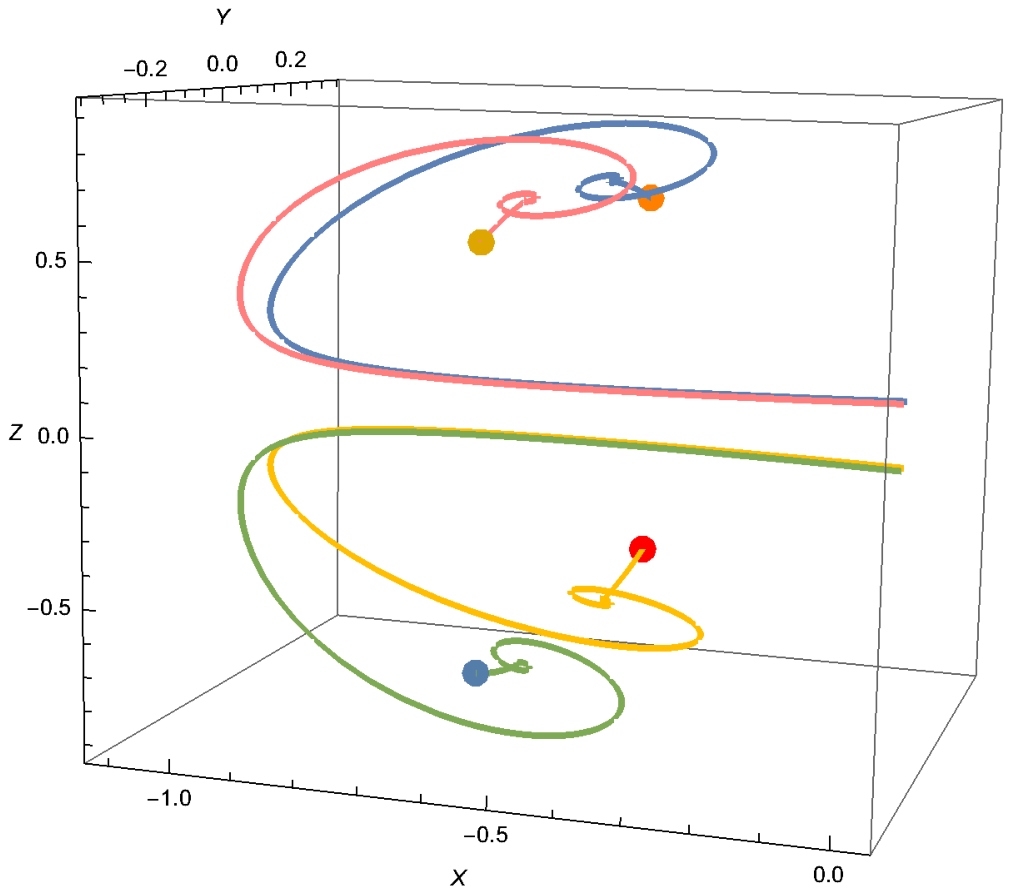}
\includegraphics[scale=0.6]{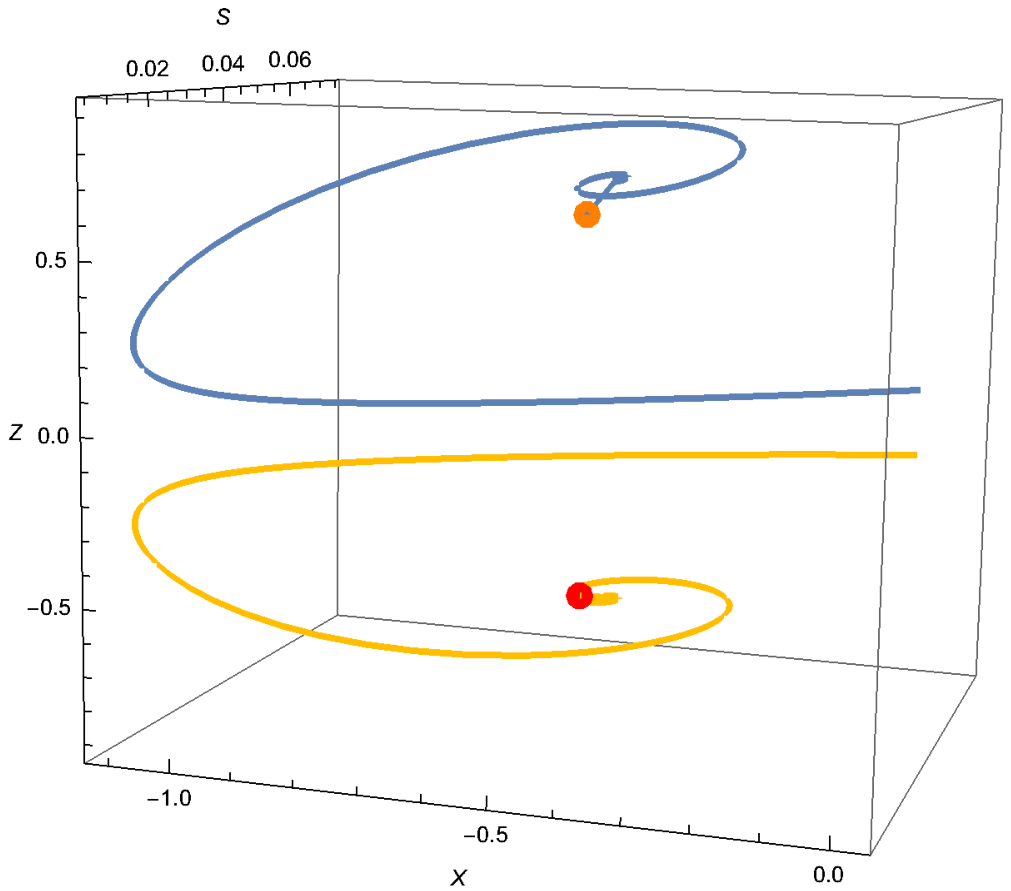}
\caption{\label{fig:++} (Left) The Phase flow in X-Y-Z space and (Right) X-S-Z space are depicted for $\lambda_{\phi}=1.2$, $\rho_{\phi}=3.3$ and $L=0.2$. The orange, red, dark yellow and dark blue circle indicate the anisotropic fixed point $(S, X, |Y|, |Z|)$, $(S, X, |Y|, -|Z|)$, $(S, X, -|Y|, |Z|)$ and $(S, X, -|Y|, -|Z|)$ respectively.}
\end{figure}

Here we give two explicit examples to study the stability of anisotropic hyperbolic inflation. The first example, we choose parameters that are not so strict in comparison with our real universe: $\{\lambda_{\phi}, \rho_{\phi}, L\}=\{1.2, 3.3, 0.2\}$. In this case, the fixed points and the eigenvalues are given by
\begin{align}
    (S, X, Y, Z)&=(0.0446429, -0.535714,\pm 0.323653,\pm 0.598947),\\
(\omega_1, \omega_2, \omega_3, \omega_4)&=
(-1.33929 + i3.72134, -1.33929 - i3.72134,\nonumber\\
&\ \ \ \ \ -2.61141, -0.0671566)
\end{align}
respectively. We can see all the real parts of eigenvalues are negative thus these fixed points are stable. In figure \ref{fig:++}, we depicted the phase flows approaching  fixed points $(S, X, |Y|, |Z|)$ in X-Y-Z space and X-S-Z space respectively. We see the trajectories converge to the anisotropic hyperbolic inflation fixed points indicated by four different circles. However, the anisotropy of this case is $\Sigma/H=0.0446429$, which is too large compared to the current data~\cite{Kim:2013gka,Naruko:2014bxa}.

As the second example, we take $\{\lambda_{\phi}, \rho_{\phi}, L\}=\{2.00000058, 333.0000000012,  0.002\}$. In this case, the fixed points and the eigenvalues are given by
\begin{align}
    (S, X, Y, Z)&=(2.93014\times 10^{-10}, -0.00598802,\pm 0.109271,\pm 0.0000513017),\\
(\omega_1, \omega_2, \omega_3, \omega_4)&=
(-1.49701 + i77.3291, -1.49701 - i77.3291,\nonumber\\
&\ \ \ \ \ -2.99401, -1.17206\times 10^{-9})
\end{align}
respectively. Again, all the real parts of the eigenvalues are negative thus the fixed points are stable. The anisotropy $\Sigma/H=2.93014\times 10^{-10}$  is consistent with observations~\cite{Kim:2013gka,Naruko:2014bxa}.

\subsection{Scaling Solution Analysis}
\begin{figure}[tpb]
\centering
\includegraphics[scale=0.5]{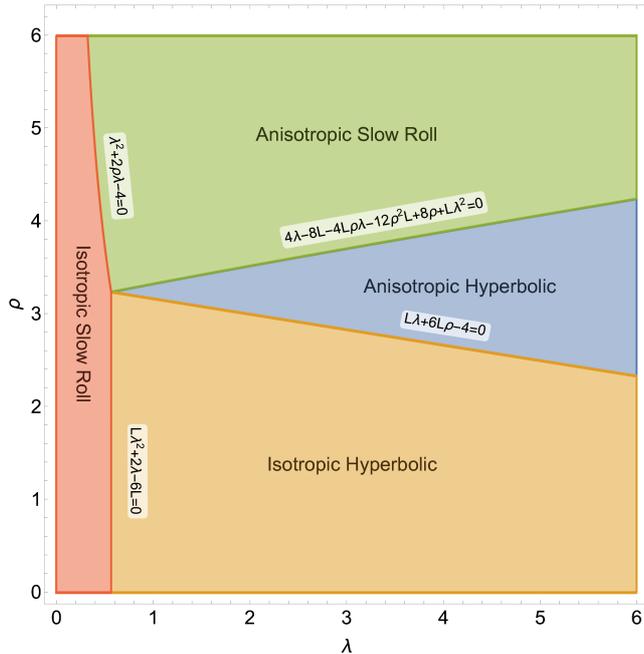}
\caption{\label{fig:PS}The parameter space of four types of inflationary solution with $L=0.2$.}
\end{figure}

We now give more discussion on the scaling solutions of this system. 
Combining all inequalities for eigenvalues in different inflationary attractors, we showed the parameter space $\lambda_{\phi}$-$\rho_{\phi}$ of four types of scaling solution dynamics in figure \ref{fig:PS} with $L=0.2$. Different types of the solutions are determined by the area where the parameters $(\lambda, \rho)$ are located. The equations of critical curves between different attractors are also shown in figure \ref{fig:PS}. There is one critical point at the intersection of the four attractors
\begin{equation}
    \left(\lambda_{*},\rho_{*}\right)=\left(-\frac{1}{L}+\sqrt{\frac{1}{L^2}+6}, \frac{5}{6L}-\frac{1}{6}\sqrt{\frac{1}{L^2}+6}\right), 
\end{equation}

\begin{figure}[tbp]
\centering
\includegraphics[scale=0.55]{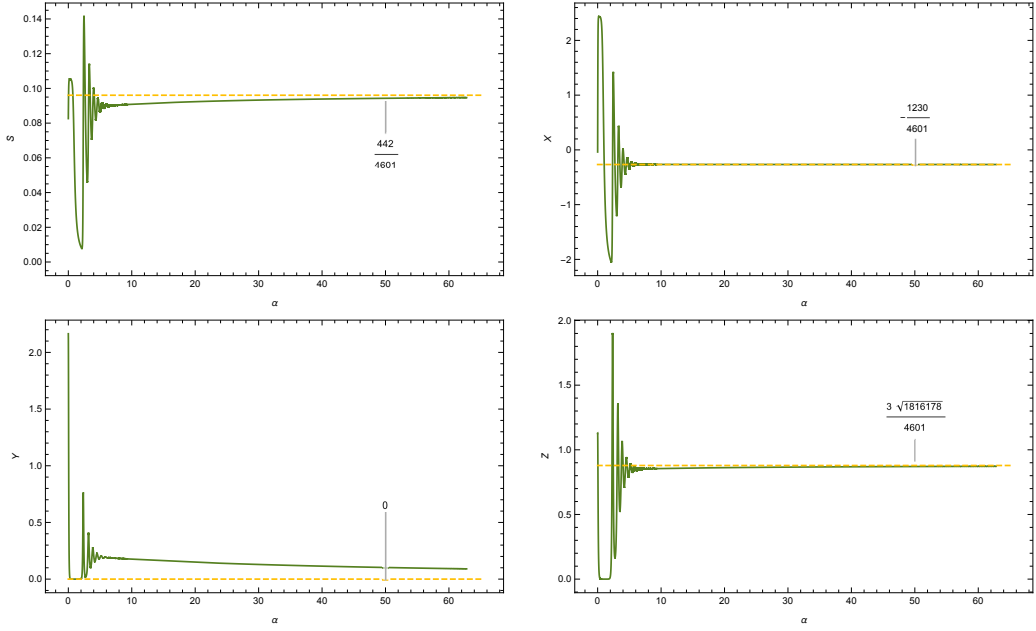}
\caption{\label{Comparision2}  Evolution of inflationary solutions (solid green curves) under exponential field space($\sim e^{\phi/L}$) for the same $\rho=7$ and $L=0.1$ as figure \ref{Comparision} but different $\lambda=2.4$. These parameters belong to area of anisotropic slow roll inflation. We also choose the same initial conditions as the figure \ref{Comparision}. The solutions converge to the scaling solutions of anisotropic slow roll inflation (\ref{AG_FP})(dashed yellow curves) rather than the anisotropic hyperbolic inflation one.}
\end{figure}

To compare with hyperbolic inflationary model in figure \ref{Comparision}, we assumed $\sinh(\phi/L)\simeq e^{\phi/L}/2$ and solved the equations of motion (\ref{eom}) numerically with the same initial conditions and $\rho_{\phi}$ but different $\lambda_{\phi}$. The $\lambda_{\phi}$ in figure \ref{Comparision2}  belongs to the anisotropic slow roll solution $\lambda<\lambda_{\text{crit}}$(the green area in figure \ref{fig:PS}) while the one in figure \ref{Comparision} belongs to the anisotropic hyperbolic solution, $\lambda>\lambda_{\text{crit}}$(the blue area in figure \ref{fig:PS}). We see the inflationary solutions $S$, $X$, $Y$ and $Z$ tend to their respective attractors finally, as we expected.

However, for a more general inflationary potential, whose parameters (\ref{pd}) depend on scalar fields, the scaling solutions may run in different attractor areas at different time. The kinetic function $f(\phi)$ can be determined by potential as $f(\phi)=e^{2C\int(V/V_{,\phi})d\phi}$, where $C$ is a constant \cite{Watanabe:2009ct}. Hence the parameter $\rho_{\phi}$ is given by
\begin{equation}
   \rho_{\phi}=\frac{2C}{\lambda_{\phi}}.\label{rhof}
\end{equation}
This is the parameter flow curve of inflation. The parameter $\rho_{\phi}$ increases (decreases) with decreasing (increasing) $\lambda_{\phi}$ for a fixed $C$. The simplest possible single-field potential is the monomial potential $V(\phi)=g\phi^n$, where $n>0$. We have $\lambda_{\phi}=n/\phi$, which is a monotonically decreasing function for $\phi>0$. If we start at a anisotropic slow roll attractor(the green area in figure \ref{fig:PS}), upon rolling down to smaller $\phi$, $\lambda_{\phi}$ increase while $\rho_{\phi}$ decrease and will hit the critical curve $4\lambda_{\phi}-8L-4\rho_{\phi} L\lambda_{\phi}-12\rho_{\phi}^2 L+8\rho_{\phi}+L\lambda_{\phi}^2=0$. Then the solutions will go into the anisotropic hyperbolic area (the blue area in figure \ref{fig:PS}).

We provide a simple toy example to illustrate this transition. We replace parameter $\lambda_{\phi}$ and $\rho_{\phi}$ in (\ref{LinearEQS})-(\ref{LinearEQZ}) with rapid transition functions near the critical curve, which is given by \cite{Christodoulidis:2019jsx}
\begin{equation}
    \lambda_{\phi}=\lambda_{0}+\Delta{\lambda}\tanh{\left[\alpha(\phi-\phi_c)\right]}
\end{equation}
arising from potential 
\begin{equation}
    V(\phi)=V_0e^{\lambda_{0}\phi}\left[\cosh{(\phi-\phi_c)}\right]^{\frac{\Delta{\lambda}}{\alpha}}.
\end{equation}
and $\rho_{\phi}$ is given by (\ref{rhof}). Such potential crosses the critical curve around $\phi=\phi_c$. For $\alpha<0$, at early times $\phi>\phi_c$ the scaling solutions are located at anisotropic slow roll area with $\lambda_{-}\equiv\lambda_{0}-\Delta{\lambda}<\lambda_{\text{crit}}$ and $\rho_{-}>\rho_{\text{crit}}$. After $\phi<\phi_c$ the solutions transition to the anisotropic hyperbolic area with $\lambda_{+}\equiv\lambda_{0}+\Delta{\lambda}>\lambda_{\text{crit}}$ and $\rho_{+}<\rho_{\text{crit}}$. In contrast, for $\alpha>0$, there is a transition from anisotropic hyperbolic inflation to anisotropic slow roll inflation around $\phi=\phi_c$. 

\begin{figure}
\centering
\includegraphics[scale=0.75]{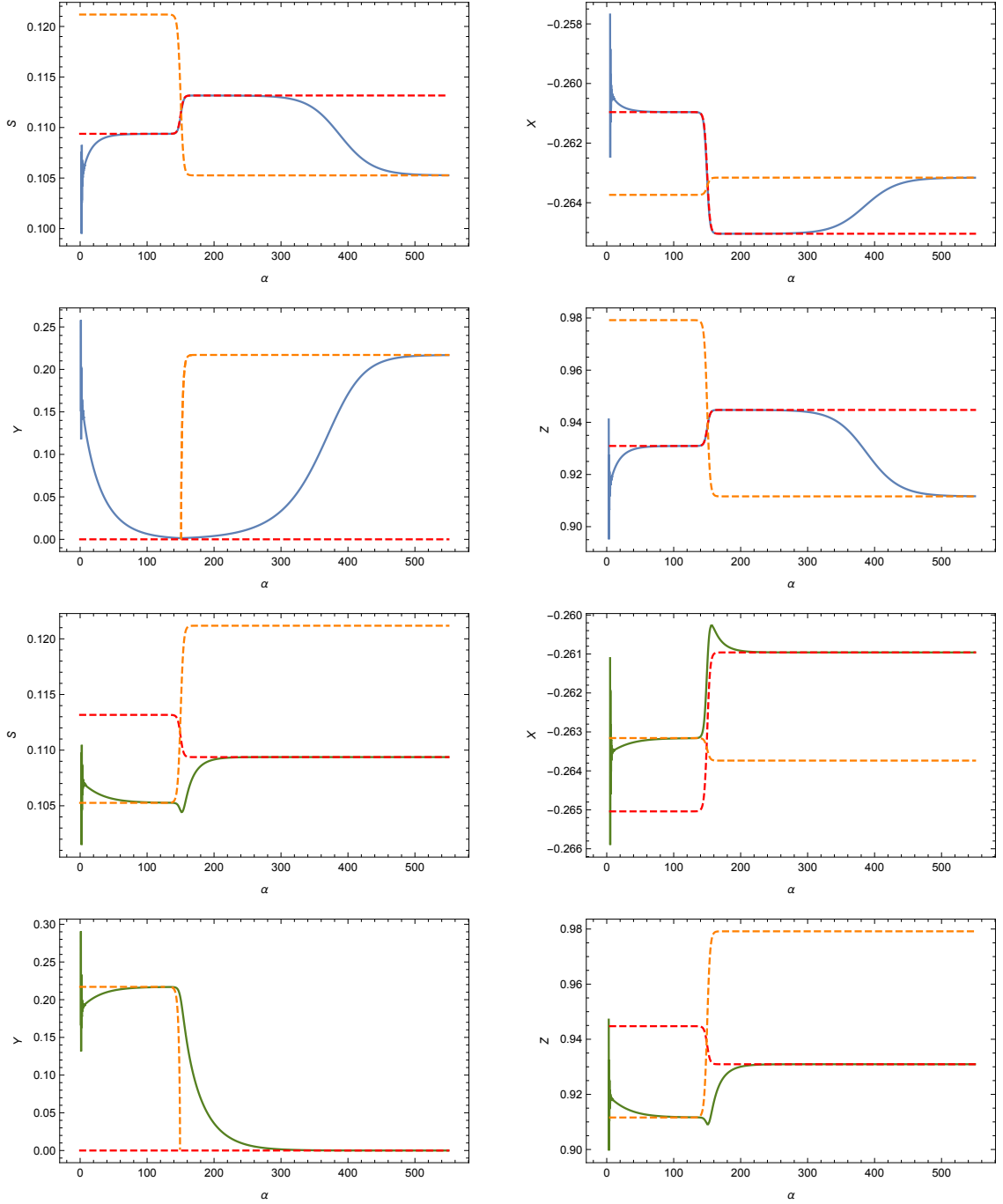}
\caption{\label{Transition} Evolution of $S$ $X$, $Y$ and $Z$(solid curves) under a transition $\lambda_{\phi}(\alpha)$. The solid blue(green) curves correspond to $\alpha=-1$($\alpha=1$). The dashed red(yellow) curves correspond to the scaling solutions (\ref{AG_FP})(solutions (\ref{AH_FP})) evaluated at $\lambda_{\phi}(\alpha)$ and $\rho_{\phi}(\alpha)=2C/\lambda_{\phi}(\alpha)$. The parameters are $L=0.1$, $C=9.8$, $\lambda_0=2.775$ and $\Delta\lambda=0.025$. The transition occurs at $\alpha_c=150$.}
\end{figure}

We assume that the solutions have converged to the scaling solution before the transition. Hence we have $\phi=X\alpha$, where $X$ is given by (\ref{AG_FP}). Figure \ref{Transition} shows an evolution of fields $S$, $X$, $Y$ and $Z$ (solid bule curves) in the case $\alpha<0$. At early times, the system follows the anisotropic slow roll inflationary scaling solution (dashed red curves). As the system evolves towards a small $\phi$, $(\lambda_{\phi}, \rho_{\phi})$ cross to the area of anisotropic hyperbolic inflation so that the solutions rapidly exit from the slow roll scaling ones. Then the system evolves along the anisotropic hyperbolic inflationay scaling solutions (yellow curves). The cases $\alpha>0$ are also shown in figure \ref{Transition} (green solid curves).

\begin{figure}[tbp]
\centering
\includegraphics[scale=0.75]{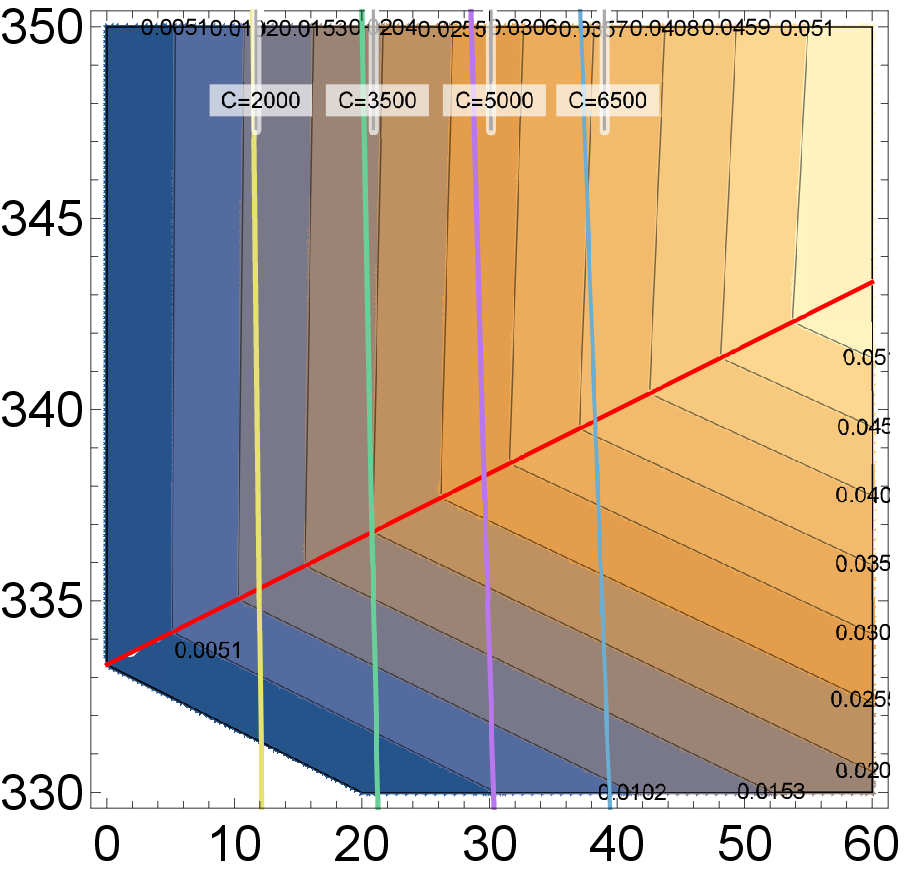}
\hspace{0.4in}
\includegraphics[scale=0.70]{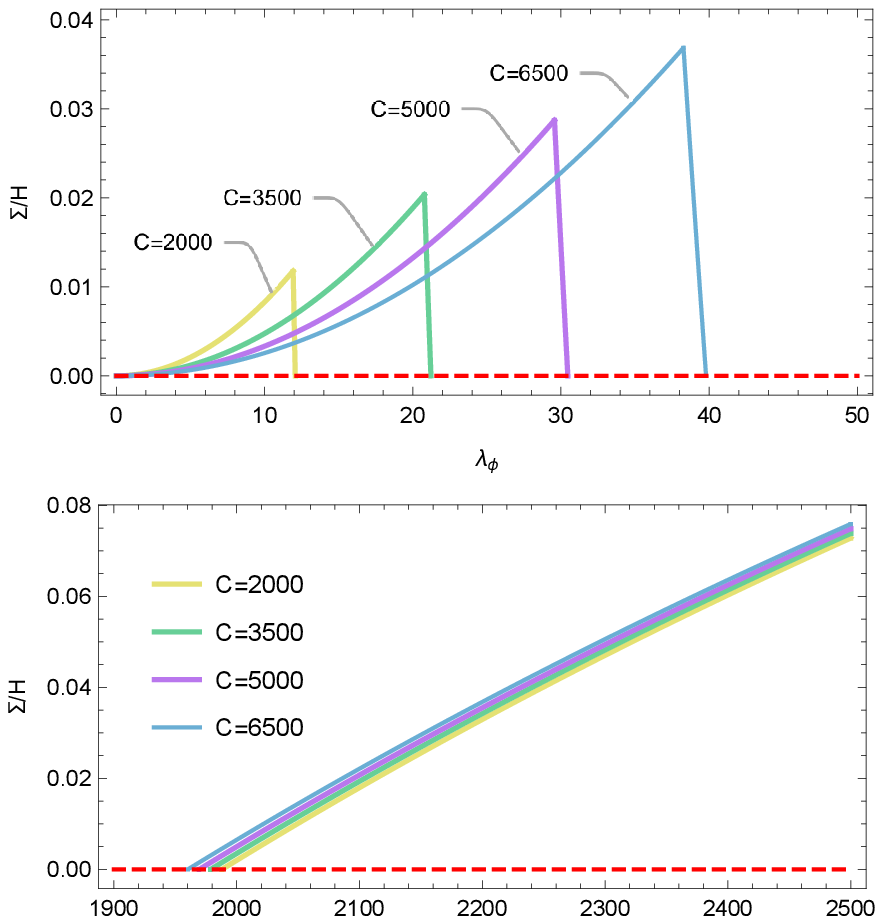}
\caption{\label{fig:contour} (Left)The contour diagram of anisotropy $\Sigma/H$ with $L=0.002$. Red curve is the critical curve between anisotropic slow roll (above) and anisotropic hyperbolic (below) inflation. The parameter flow curves with different $C$ are also shown (yellow, green, purple and white curves). The closer to the critical curve the larger anisotropy $\Sigma/H$(right).}
\end{figure}

We also analyze the evolution of anisotropy $\Sigma/H$ in these two ansiotropic areas. $\Sigma/H$ is a continuous function, but its first order derivative is not at the critical curve. After substituting the parameter flow equation (\ref{rhof}) into $\Sigma/H$, we obtain
\begin{equation}
\frac{\Sigma}{H}=\left\{
\begin{aligned}
\frac{2\lambda_{\phi}^2\left(\lambda_{\phi}^2+4C-4\right)}{\lambda_{\phi}^4+\left(16C+8\right)\lambda_{\phi}^2+48C^2}, \ \ \ \ \ &\text{Anisotropic Slow Roll}\\
\frac{L\lambda_{\phi}^2-4\lambda_{\phi}+12LC}{2\lambda_{\phi}\left(L\lambda_{\phi}+2\right)}\ \ \ \ \ ,\ \ \ \ \ &\text{Anisotropic Hyperbolic}
\end{aligned}
\right.
\end{equation}
where $C$ is the parameter of kinetic function. We show the contour diagram of the anisotropy in figure \ref{fig:contour}. To compare with inflation, we chose small $L=0.002$ and consider the area where $\rho_{\phi}\gg\lambda_{\phi}$. For a fixed $\lambda_{\phi}$, the anisotropy becomes larger as $\rho_{\phi}$ decreases before hitting the critical curve (red curve). After crossing the critical curve, $\Sigma/H$ becomes smaller as $\rho_{\phi}$ decreases.
The parameter flow curves (\ref{rhof}) with different $C$ are shown in figure \ref{fig:contour}. Note that for inflationary model regime $\rho_{\phi}\gg\lambda_{\phi}$, there should be $C\gg\lambda_{\phi}^2/2$. We see the maximum of $\Sigma/H$ is at the critical curve and the anisotropy becomes smaller rapidly with increasing $\lambda_{\phi}$ after crossing the curve. The parameter flow curves soon cross to the isotropic hyperbolic inflationary area thus $\Sigma/H$ drops to zero. 
Afterward, the parameter flow curves will cross to the anisotropic hyperbolic inflationary area once again thus $\Sigma/H$ becomes larger again when $\lambda_{\phi}$ is large enough (see figure \ref{fig:contour}). However, at this time $C\lesssim\lambda_{\phi}^2/2$ thus  inflation has already ended.

\section{Conclusion}
We argued the importance of understanding the phase space structure of inflationary dynamics
in the context of multi-scalar-gauge-field models. 
In this paper, as a first step, we have studied the effect of a gauge field on hyperbolic inflation. In our model, there are one massive radial field $\phi$ and one massless angular field $\theta$, where $\phi$ is coupled to a $U(1)$ gauge field with a gauge kinetic function. We considered an exponential potential and an exponential kinetic function of radial field $\phi$. For $\phi\gg L$ regime where geometry of field-space is described by the metric $\simeq e^{\phi/L}/2$, there exist analytical power-law solutions. These solutions include the anisotropic hyperbolic inflation attractors, not only the original slow-roll ones. We showed that the anisotropy is proportional to the slow-roll parameter, and is suppressed by small $L$. We also analyzed the stability of this dynamical system and found the attracotrs are stable in the parameter area of anisotropic hyperbolic inflation. We also discussed more general potentials where trajectories run 
in the parameter space $(\lambda, \rho)$. We analyzed the dynamics of these scaling solutions and found when following the parameter flow curve, the anisotropy approaches to the maximum at the critical curve between anisotropic slow roll inflation and anisotropic hyperbolic inflation.
From our results, we can argue that destabilization of the conventional 
slow roll inflation gives rise to a fertile phase space structure 
in the multi-scalar-gauge-field models.

We found both the expansion of universe and the slow-roll parameter only depends on the $L\lambda$. This is because the angular field $\theta$ do not couple directly to the $U(1)$ gauge field so that the expansion can be solved from the equation of motion of $\theta$. One can regard the field $\theta$ as axion fields, which are a classes of pseudo-scalar fields motivated in particle physics and string theory (see \cite{Marsh:2015xka,Pajer:2013fsa} for reviews). Then one should introduce the Chern-Simon coupling $\sim\theta F \tilde{F}$ in the model. However, for homogeneous backgrounds this term do not contribute to the local equations of motion. To concern this coupling, one should consider the non-abelian case, e.g. $SU(2)$ gauge fields so that there is a Yang-Mills coupling term in the equations \cite{Murata:2011wv,Maleknejad:2011jw,Maleknejad:2011sq,Adshead:2012kp,Adshead:2012qe,Obata:2014loa,Maeda:2012eg,Maeda:2013daa,Adshead:2018emn,Gao:2021qwl}. In some models, the anisotropy can survive during inflationary period (for example, \cite{Murata:2011wv,Adshead:2018emn,Gao:2021qwl}). On the other hand, the axion coupled with gauge fields $SU(2)$, i.e., the so-called Chromo-natural inflationary model under a anisotropic background has been studied recently. It showed that the anisotropy can not survive during inflation and the system has stable isotropic attractors for a wide range of parameters space \cite{Maleknejad:2013npa,Wolfson:2020fqz,Wolfson:2021fya}. Hence, it's worth to extend our model to the non-Abelian case. the One can also consider a massive angular field that potential is depended on $\phi$ and $\theta$. Moreover, It is also interesting to compute the perturbations of this model to compare with observations.
We should also study general multi-scalar-gauge-field inflation from the point of destabilization of the conventional slow roll inflation.


\appendix
\section{Stability of Anisotropic Hyperbolic Inflationary Fixed Point}\label{AppA}
To analyze the linear stability of fixed points, one can obtain the linear equations of perturbations of variables, $d\delta X^a/d\alpha=J^a_{\ b}\delta X^b$. The stability can be analyzed by $\delta X^a=e^{\omega\alpha}\delta\tilde{X}$, where $\omega$ are eigenvalues of the Jacobian matrix $J^a_{\ b}$. If the real parts of the eigenvalues are negative, the fixed point is stable.

\begin{figure}[htbp]
\centering
\includegraphics[scale=0.9]{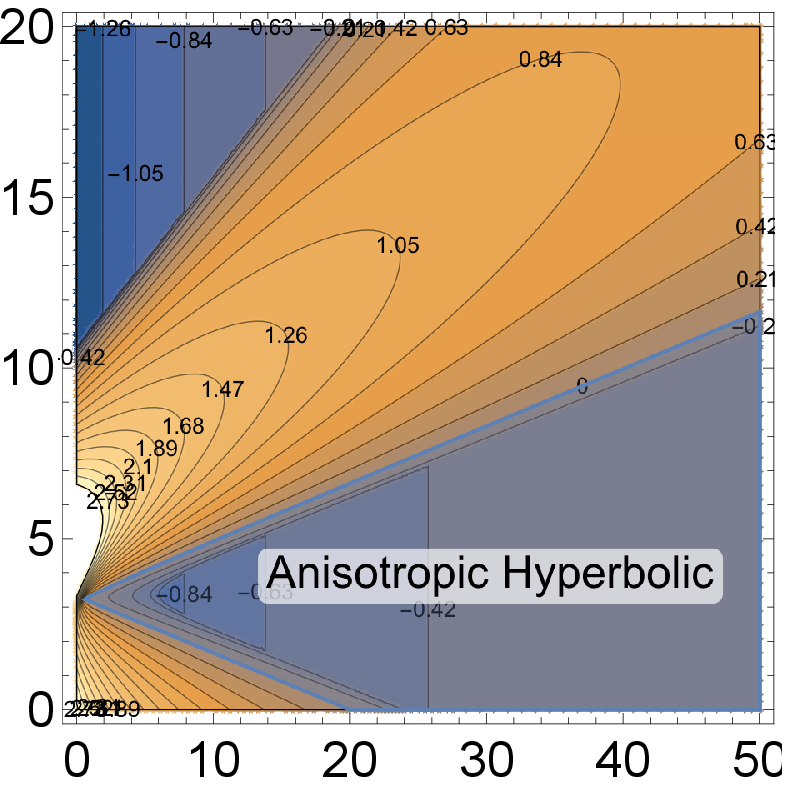}
\ \ \ \ 
\includegraphics[scale=0.9]{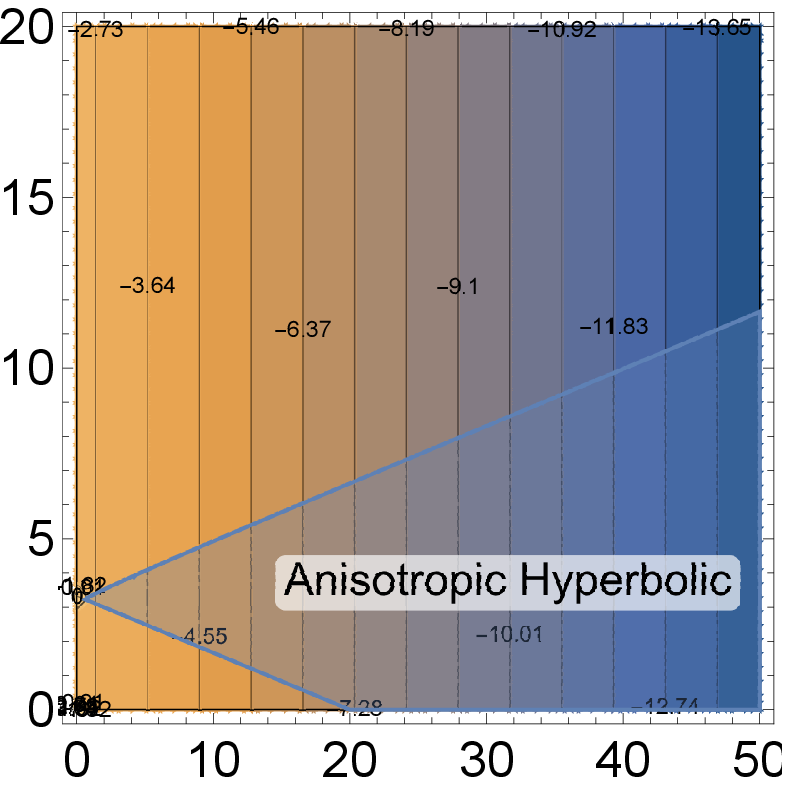}
\caption{\label{fig:eigenvalue} The parameter space of real parts of $\omega_2$(left) and $\omega_4$(right) with $L=0.2$. The areas of anisotropic inflation (bottom right blue area) cover both the negative parts of them.}
\end{figure}

Here we provide the full expressions of eigenvalue of Jacobian matrix of anisotropic hyperbolic inflation perturbtions. The eigenvalues are given by
\begin{eqnarray}
    \omega_{1,2}&=&-\frac{3}{\lambda  L+2}\mp\frac{3\sqrt{A(\lambda,\rho)}}{2 L (\lambda  L+2)^2},\ \ \ \ \ \   \omega_{3,4}=-\frac{3}{\lambda  L+2}\mp\frac{3\sqrt{B(\lambda,\rho)}}{2 L (\lambda  L+2)^2},
\end{eqnarray}
where $A$ and $B$ are given by
\begin{align}
    A&=(L^2 (\lambda +2 \rho ) (\lambda  L+2)^4 (64 (\lambda +2
   \rho )+L^4 (8 \lambda ^3 (3 \rho ^2-2)-16 \lambda ^2 \rho  (\rho
   ^2-6)\nonumber\\
   &\ \ \ -6 \lambda ^4 \rho+\lambda ^5+\lambda  (80-48 \rho ^4)+96 \rho 
   (3 \rho ^4-1))+4 L^3 (4 \lambda ^2 (8 \rho ^2-9)-16
   \lambda ^3 \rho +3 \lambda ^4\nonumber\\
   &\ \ \ +16 \lambda  \rho  (2 \rho ^2+5)-240 \rho
   ^4+48 \rho ^2+64)+4 L^2 (-50 \lambda ^2 \rho +13 \lambda ^3-4 \lambda 
   (\rho ^2+20)\nonumber\\
   &\ \ \ +8 \rho  (37 \rho ^2-4))+32 L (3 \lambda -10
   \rho ) (\lambda +2 \rho )))^{1/2}\nonumber\\
   &\ \ \ +L (\lambda  L+2)^2 (L (-6 \lambda ^2+8
   \lambda  \rho-L (-2 \lambda ^2 \rho +\lambda ^3+4 \lambda  (\rho
   ^2-1)+24 (\rho ^3+\rho ))\nonumber\\
   &\ \ \ +40 \rho ^2+36)-8 (\lambda +2
   \rho )),\label{eigenvalueA}
\\\nonumber\\
   B&=L (\lambda  L+2)^2 (L (-6 \lambda ^2+8 \lambda  \rho -L (-2 \lambda
   ^2 \rho +\lambda ^3+4 \lambda (\rho ^2-1)+24 (\rho ^3+\rho
   ))\nonumber\\
   &\ \ \ +40 \rho ^2+36)-8 (\lambda +2 \rho ))-(L^2 (\lambda +2
   \rho ) (\lambda  L+2)^4 (64 (\lambda +2 \rho )\nonumber\\
   &\ \ \ +L^4 (8 \lambda ^3 (3
   \rho ^2-2)-16 \lambda ^2 \rho  (\rho ^2-6)-6 \lambda ^4 \rho
   +\lambda ^5+\lambda  (80-48 \rho ^4)+96 \rho  (3 \rho
   ^4-1))\nonumber\\
   &\ \ \ +4 L^3 (4 \lambda ^2 (8 \rho ^2-9)-16 \lambda ^3
   \rho +3 \lambda ^4+16 \lambda  \rho  (2 \rho ^2+5)-240 \rho ^4+48 \rho
   ^2+64)\nonumber\\
   &\ \ \ +4 L^2 (-50 \lambda ^2 \rho +13 \lambda ^3-4 \lambda  (\rho
   ^2+20)+8 \rho  (37 \rho ^2-4))\nonumber\\
   &\ \ \ +32 L (3 \lambda -10 \rho )
   (\lambda +2 \rho )))^{1/2},\label{eigenvalueB}
\end{align}
It's obvious that $\omega_1,\omega_3<0$. For $\omega_2$ and $\omega_4$, we show in figure \ref{fig:eigenvalue} the real parts of these two eigenvalues in $\lambda$-$\rho$ plane. We found the area of anisotropic hyperbolic inflation covers the areas of negative real parts of $\omega_2$ and $\omega_4$. Therefore the anisotropic hyperbolic inflation solutions are stable.

\acknowledgments

J.\,S. was in part supported by JSPS KAKENHI Grant Numbers JP17H02894, JP17K18778, JP20H01902.
C-B.\,C. was supported by Japanese Government (MEXT) Scholarship and China Scholarship Council (CSC).



\begin{thebibliography}{99}
\bibitem{Guth:1980zm}
A.~H.~Guth,
``The Inflationary Universe: A Possible Solution to the Horizon and Flatness Problems,''
Phys. Rev. D \textbf{23}, 347-356 (1981)

\bibitem{Sato:1981ds}
K.~Sato,
``Cosmological Baryon Number Domain Structure and the First Order Phase Transition of a Vacuum,''
Phys. Lett. B \textbf{99}, 66-70 (1981)

\bibitem{Linde:1981mu}
A.~D.~Linde,
``A New Inflationary Universe Scenario: A Possible Solution of the Horizon, Flatness, Homogeneity, Isotropy and Primordial Monopole Problems,''
Phys. Lett. B \textbf{108}, 389-393 (1982)

\bibitem{Albrecht:1982wi}
A.~Albrecht and P.~J.~Steinhardt,
``Cosmology for Grand Unified Theories with Radiatively Induced Symmetry Breaking,''
Phys. Rev. Lett. \textbf{48}, 1220-1223 (1982)

\bibitem{Ade:2015lrj} 
  P.~A.~R.~Ade {\it et al.} [Planck Collaboration],
  ``Planck 2015 results. XX. Constraints on inflation,''
  Astron.\ Astrophys.\  {\bf 594}, A20 (2016)
  [arXiv:1502.02114 [astro-ph.CO]].
  
\bibitem{Array:2015xqh}
P.~A.~R.~Ade \textit{et al.} [BICEP2 and Keck Array],
``Improved Constraints on Cosmology and Foregrounds from BICEP2 and Keck Array Cosmic Microwave Background Data with Inclusion of 95 GHz Band,''
Phys. Rev. Lett. \textbf{116}, 031302 (2016)
[arXiv:1510.09217 [astro-ph.CO]].

\bibitem{Ackerman:2007nb}
L.~Ackerman, S.~M.~Carroll and M.~B.~Wise,
``Imprints of a Primordial Preferred Direction on the Microwave Background,''
Phys. Rev. D \textbf{75}, 083502 (2007)
[erratum: Phys. Rev. D \textbf{80}, 069901 (2009)]
doi:10.1103/PhysRevD.75.083502
[arXiv:astro-ph/0701357 [astro-ph]].

\bibitem{Planck:2015igc}
P.~A.~R.~Ade \textit{et al.} [Planck],
``Planck 2015 results. XVI. Isotropy and statistics of the CMB,''
Astron. Astrophys. \textbf{594}, A16 (2016)
doi:10.1051/0004-6361/201526681
[arXiv:1506.07135 [astro-ph.CO]].

\bibitem{Watanabe:2009ct}
M.~a.~Watanabe, S.~Kanno and J.~Soda,
``Inflationary Universe with Anisotropic Hair,''
Phys. Rev. Lett. \textbf{102}, 191302 (2009)
[arXiv:0902.2833 [hep-th]].

\bibitem{Kanno:2009ei}
S.~Kanno, J.~Soda and M.~a.~Watanabe,
``Cosmological Magnetic Fields from Inflation and Backreaction,''
JCAP \textbf{12}, 009 (2009)
[arXiv:0908.3509 [astro-ph.CO]].

\bibitem{Soda:2012zm}
J.~Soda,
``Statistical Anisotropy from Anisotropic Inflation,''
Class. Quant. Grav. \textbf{29}, 083001 (2012)
[arXiv:1201.6434 [hep-th]].

\bibitem{Maleknejad:2012fw}
A.~Maleknejad, M.~M.~Sheikh-Jabbari and J.~Soda,
``Gauge Fields and Inflation,''
Phys. Rept. \textbf{528}, 161-261 (2013)
[arXiv:1212.2921 [hep-th]].

\bibitem{Vafa:2005ui}
C.~Vafa,
``The String landscape and the swampland,''
[arXiv:hep-th/0509212 [hep-th]].

\bibitem{Ooguri:2006in}
H.~Ooguri and C.~Vafa,
``On the Geometry of the String Landscape and the Swampland,''
Nucl. Phys. B \textbf{766}, 21-33 (2007)
doi:10.1016/j.nuclphysb.2006.10.033
[arXiv:hep-th/0605264 [hep-th]].

\bibitem{Obied:2018sgi}
G.~Obied, H.~Ooguri, L.~Spodyneiko and C.~Vafa,
``De Sitter Space and the Swampland,''
[arXiv:1806.08362 [hep-th]].

\bibitem{Agrawal:2018own}
P.~Agrawal, G.~Obied, P.~J.~Steinhardt and C.~Vafa,
``On the Cosmological Implications of the String Swampland,''
Phys. Lett. B \textbf{784}, 271-276 (2018)
[arXiv:1806.09718 [hep-th]].

\bibitem{Garg:2018reu}
S.~K.~Garg and C.~Krishnan,
``Bounds on Slow Roll and the de Sitter Swampland,''
JHEP \textbf{11}, 075 (2019)
[arXiv:1807.05193 [hep-th]].

\bibitem{Renaux-Petel:2015mga}
S.~Renaux-Petel and K.~Turzy\'nski,
``Geometrical Destabilization of Inflation,''
Phys. Rev. Lett. \textbf{117}, no.14, 141301 (2016)
[arXiv:1510.01281 [astro-ph.CO]].

\bibitem{Easson:2007dh}
D.~A.~Easson, R.~Gregory, D.~F.~Mota, G.~Tasinato and I.~Zavala,
``Spinflation,''
JCAP \textbf{02}, 010 (2008)
[arXiv:0709.2666 [hep-th]].

\bibitem{Brown:2017osf}
A.~R.~Brown,
``Hyperbolic Inflation,''
Phys. Rev. Lett. \textbf{121}, no.25, 251601 (2018)
[arXiv:1705.03023 [hep-th]].

\bibitem{Mizuno:2017idt}
S.~Mizuno and S.~Mukohyama,
``Primordial perturbations from inflation with a hyperbolic field-space,''
Phys. Rev. D \textbf{96}, no.10, 103533 (2017)
[arXiv:1707.05125 [hep-th]].

\bibitem{Bounakis:2020xaw}
M.~Bounakis, I.~G.~Moss and G.~Rigopoulos,
``Observational constraints on Hyperinflation,''
JCAP \textbf{02}, 006 (2021)
[arXiv:2010.06461 [gr-qc]].

\bibitem{Bjorkmo:2019aev}
T.~Bjorkmo and M.~C.~D.~Marsh,
``Hyperinflation generalised: from its attractor mechanism to its tension with the \textquoteleft{}swampland conditions\textquoteright{},''
JHEP \textbf{04}, 172 (2019)
[arXiv:1901.08603 [hep-th]].

\bibitem{Ohashi:2013mka}
J.~Ohashi, J.~Soda and S.~Tsujikawa,
``Anisotropic Non-Gaussianity from a Two-Form Field,''
Phys. Rev. D \textbf{87}, no.8, 083520 (2013)
[arXiv:1303.7340 [astro-ph.CO]].

\bibitem{Ito:2015sxj}
A.~Ito and J.~Soda,
``Designing Anisotropic Inflation with Form Fields,''
Phys. Rev. D \textbf{92}, no.12, 123533 (2015)
[arXiv:1506.02450 [hep-th]].

\bibitem{Halliwell:1986ja}
J.~J.~Halliwell,
``Scalar Fields in Cosmology with an Exponential Potential,''
Phys. Lett. B \textbf{185}, 341 (1987).

\bibitem{Copeland:1997et}
E.~J.~Copeland, A.~R.~Liddle and D.~Wands,
``Exponential potentials and cosmological scaling solutions,''
Phys. Rev. D \textbf{57}, 4686-4690 (1998)
[arXiv:gr-qc/9711068 [gr-qc]].

\bibitem{Kanno:2010nr}
S.~Kanno, J.~Soda and M.~a.~Watanabe,
``Anisotropic Power-law Inflation,''
JCAP \textbf{12}, 024 (2010)
[arXiv:1010.5307 [hep-th]].

\bibitem{Yamamoto:2012tq}
K.~Yamamoto, M.~a.~Watanabe and J.~Soda,
``Inflation with Multi-Vector-Hair: The Fate of Anisotropy,''
Class. Quant. Grav. \textbf{29}, 145008 (2012)
[arXiv:1201.5309 [hep-th]].


\bibitem{Lahiri:2016jqv}
S.~Lahiri,
``Anisotropic inflation in Gauss-Bonnet gravity,''
JCAP \textbf{09}, 025 (2016)
doi:10.1088/1475-7516/2016/09/025
[arXiv:1605.09247 [hep-th]].

\bibitem{Ohashi:2013pca}
J.~Ohashi, J.~Soda and S.~Tsujikawa,
``Anisotropic power-law k-inflation,''
Phys. Rev. D \textbf{88}, 103517 (2013)
doi:10.1103/PhysRevD.88.103517
[arXiv:1310.3053 [hep-th]].

\bibitem{Ito:2017bnn}
A.~Ito and J.~Soda,
``Anisotropic Constant-roll Inflation,''
Eur. Phys. J. C \textbf{78}, no.1, 55 (2018)
[arXiv:1710.09701 [hep-th]].

\bibitem{Holland:2017cza}
J.~Holland, S.~Kanno and I.~Zavala,
``Anisotropic Inflation with Derivative Couplings,''
Phys. Rev. D \textbf{97}, no.10, 103534 (2018)
[arXiv:1711.07450 [hep-th]].

\bibitem{Do:2020hjf}
T.~Q.~Do,
``Stable small spatial hairs in a power-law k-inflation model,''
Eur. Phys. J. C \textbf{81}, no.1, 77 (2021)
[arXiv:2007.04867 [gr-qc]].

\bibitem{Do:2021lyf}
T.~Q.~Do and W.~F.~Kao,
``Anisotropic power-law inflation for a model of two scalar and two vector fields,''
[arXiv:2104.14100 [gr-qc]].

\bibitem{Christodoulidis:2019jsx}
P.~Christodoulidis, D.~Roest and E.~I.~Sfakianakis,
``Scaling attractors in multi-field inflation,''
JCAP \textbf{12}, 059 (2019)
[arXiv:1903.06116 [hep-th]].

\bibitem{Christodoulidis:2018msl}
P.~Christodoulidis,
``Probing the inflationary evolution using analytical solutions,''
[arXiv:1811.06456 [astro-ph.CO]].

\bibitem{Kim:2013gka}
J.~Kim and E.~Komatsu,
``Limits on anisotropic inflation from the Planck data,''
Phys. Rev. D \textbf{88}, 101301 (2013)
[arXiv:1310.1605 [astro-ph.CO]].

\bibitem{Naruko:2014bxa}
A.~Naruko, E.~Komatsu and M.~Yamaguchi,
``Anisotropic inflation reexamined: upper bound on broken rotational invariance during inflation,''
JCAP \textbf{04}, 045 (2015)
[arXiv:1411.5489 [astro-ph.CO]].

\bibitem{Wald:1983ky}
R.~M.~Wald,
``Asymptotic behavior of homogeneous cosmological models in the presence of a positive cosmological constant,''
Phys. Rev. D \textbf{28}, 2118-2120 (1983).

\bibitem{Soda:2014awa}
J.~Soda,
``Anisotropic Power-law Inflation:A counter example to the cosmic no-hair conjecture,''
J. Phys. Conf. Ser. \textbf{600}, no.1, 012026 (2015)
[arXiv:1410.8643 [gr-qc]].

\bibitem{Marsh:2015xka}
D.~J.~E.~Marsh,
``Axion Cosmology,''
Phys. Rept. \textbf{643}, 1-79 (2016)
[arXiv:1510.07633 [astro-ph.CO]].

\bibitem{Pajer:2013fsa}
E.~Pajer and M.~Peloso,
``A review of Axion Inflation in the era of Planck,''
Class. Quant. Grav. \textbf{30}, 214002 (2013)
[arXiv:1305.3557 [hep-th]].

\bibitem{Murata:2011wv}
K.~Murata and J.~Soda,
``Anisotropic Inflation with Non-Abelian Gauge Kinetic Function,''
JCAP \textbf{06}, 037 (2011)
[arXiv:1103.6164 [hep-th]].

\bibitem{Maleknejad:2011jw}
A.~Maleknejad and M.~M.~Sheikh-Jabbari,
``Gauge-flation: Inflation From Non-Abelian Gauge Fields,''
Phys. Lett. B \textbf{723}, 224-228 (2013)
[arXiv:1102.1513 [hep-ph]].

\bibitem{Maleknejad:2011sq}
A.~Maleknejad and M.~M.~Sheikh-Jabbari,
``Non-Abelian Gauge Field Inflation,''
Phys. Rev. D \textbf{84}, 043515 (2011)
[arXiv:1102.1932 [hep-ph]].

\bibitem{Adshead:2012kp}
P.~Adshead and M.~Wyman,
``Chromo-Natural Inflation: Natural inflation on a steep potential with classical non-Abelian gauge fields,''
Phys. Rev. Lett. \textbf{108}, 261302 (2012)
doi:10.1103/PhysRevLett.108.261302
[arXiv:1202.2366 [hep-th]].

\bibitem{Adshead:2012qe}
P.~Adshead and M.~Wyman,
``Gauge-flation trajectories in Chromo-Natural Inflation,''
Phys. Rev. D \textbf{86}, 043530 (2012)
[arXiv:1203.2264 [hep-th]].

\bibitem{Obata:2014loa}
I.~Obata, T.~Miura and J.~Soda,
``Chromo-Natural Inflation in the Axiverse,''
Phys. Rev. D \textbf{92}, no.6, 063516 (2015)
[arXiv:1412.7620 [hep-ph]].

\bibitem{Maeda:2012eg}
K.~i.~Maeda and K.~Yamamoto,
``Inflationary Dynamics with a Non-Abelian Gauge Field,''
Phys. Rev. D \textbf{87}, no.2, 023528 (2013)
[arXiv:1210.4054 [astro-ph.CO]].

\bibitem{Maeda:2013daa}
K.~i.~Maeda and K.~Yamamoto,
``Stability analysis of inflation with an SU(2) gauge field,''
JCAP \textbf{12}, 018 (2013)
[arXiv:1310.6916 [gr-qc]].

\bibitem{Adshead:2018emn}
P.~Adshead and A.~Liu,
``Anisotropic Massive Gauge-flation,''
JCAP \textbf{07}, 052 (2018)
doi:10.1088/1475-7516/2018/07/052
[arXiv:1803.07168 [astro-ph.CO]].

\bibitem{Gao:2021qwl}
P.~Gao, K.~Takahashi, A.~Ito and J.~Soda,
``Cosmic No-hair Conjecture and Inflation with an SU(3) Gauge Field,''
[arXiv:2107.00264 [hep-th]].

\bibitem{Maleknejad:2013npa}
A.~Maleknejad and E.~Erfani,
``Chromo-Natural Model in Anisotropic Background,''
JCAP \textbf{03}, 016 (2014)
doi:10.1088/1475-7516/2014/03/016
[arXiv:1311.3361 [hep-th]].

\bibitem{Wolfson:2020fqz}
I.~Wolfson, A.~Maleknejad and E.~Komatsu,
``How attractive is the isotropic attractor solution of axion-SU(2) inflation?,''
JCAP \textbf{09}, 047 (2020)
doi:10.1088/1475-7516/2020/09/047
[arXiv:2003.01617 [gr-qc]].

\bibitem{Wolfson:2021fya}
I.~Wolfson, A.~Maleknejad, T.~Murata, E.~Komatsu and T.~Kobayashi,
``The isotropic attractor solution of axion-SU(2) inflation: Universal isotropization in Bianchi type-I geometry,''
[arXiv:2105.06259 [gr-qc]].





\end{thebibliography}
\end{document}